\documentclass[conference]{IEEEtran}

\usepackage{cite}
\usepackage{amsmath,amssymb,amsfonts}
\usepackage{algorithmic}
\usepackage{graphicx}
\usepackage{textcomp}
\usepackage{xcolor}
\usepackage{booktabs}
\usepackage{listings}
\usepackage{float}

\usepackage{xcolor} % For easy color management
\newcommand{\editcolor}{black} % Define the edit color here
\newcommand{\edit}[1]{{\color{\editcolor} #1}} % Command to color your edits

\begin{document}

\title{Latency-Neutral Passive Cyber Monitoring for Grid-Edge Operational Technology Networks}

\author{\IEEEauthorblockN{Damian Byrne, Emad Abukhousa, Jason Reed, and Saman Zonouz}
\IEEEauthorblockA{\textit{School of Electrical and Computer Engineering, Georgia Institute of Technology, Atlanta, USA}\\
\{dbyrne31, emadak, jreed70, szonouz6\}@gatech.edu}
}

\maketitle

\begin{abstract}
Deterministic industrial networks often present an operational trade-off between maintaining microsecond timing of primary control loops and capturing security telemetry. This work investigates the engineering feasibility of the Operational Technology Supervisor (OTSV), a three-tier passive monitoring framework designed to resolve this visibility challenge within timing-sensitive grid-edge infrastructure. \edit{By utilizing a Layer-3-unconfigured passive ingestion mode on bare-metal hardware via a physical switched port analyzer (SPAN) loop, the deep packet inspection engine is isolated from the production network, implementing a software-enforced unidirectional export policy.} Performance validation conducted over a continuous audit comprising 32,927 active-monitoring round-trip time (RTT) samples yielded stable latency distributions across a laboratory power-monitoring loop running Modbus TCP and IEC 61850 MMS protocols. \edit{The measured latency delta between the unmonitored baseline
and the active monitoring state was less than 0.12 milliseconds,
supporting the feasibility of continuous high-fidelity security
observation alongside stringent grid-edge metering timing bounds
without measurably degrading the monitored reporting path.}
\end{abstract}


\section{Introduction}
The systematic dissolution of the physical air gap in critical infrastructure has fundamentally altered the threat profile of industrial facilities \cite{nist80082r3, bodenheim2014}. Modern hyperscale data centers and critical power facilities rely on the deep integration of Information Technology (IT) and Operational Technology (OT) to optimize mission-critical systems at the grid edge, such as computer room air conditioning arrays, power quality meters (PQMs), and power distribution units \cite{shehabi2024}. While this interoperability facilitates real-time telemetry and predictive maintenance, it forces legacy firmware, which often lacks native cryptographic protections, onto the exposed network boundary \cite{iec62351}. This exposure creates a substantial attack surface where lateral movement from compromised IT assets can directly impact physical system integrity.

Traditional cybersecurity solutions require continuous deep packet inspection (DPI) and automated asset discovery to maintain an active defense posture \cite{formby2016, wedgbury2015}. However, the deterministic physics of OT networks complicates the deployment of standard security tooling \cite{pospisil2018}. Unlike IT traffic where asynchronous packet delivery is standard, OT environments demand isochronous performance to maintain control loop synchronization. Standard active reconnaissance—characterized by aggressive transmission control protocol (TCP) handshakes and service probes—is often incompatible with sensitive OT architectures, particularly legacy programmable logic controller (PLC) stacks \cite{niedermaier2018woot, niedermaier2018propfuzz}. Even marginal deterministic jitter induced by active security prodding can desynchronize proportional-integral-derivative control loops, lead to network congestion or immediate controller lock-states, and potentially invoke buffer overflows \cite{pospisil2018}. This conflict represents the Visibility-Stability Paradox: the exact telemetry required to secure the infrastructure poses a direct risk to operational uptime.

This work proposes the OT Supervisor (OTSV) architecture as a non-intrusive solution to this paradox. Positioned as a deployable solution for converged IT/OT environments managing facility power monitoring, power quality meters, and AI-driven grid-edge load configurations, this paper presents an empirical latency-neutrality audit of a passive monitoring stack under attack-concurrent deep packet inspection load. Rather than inventing passive monitoring, this study contributes a rigorous experimental feasibility evaluation of a decoupled monitoring design across real power-monitoring hardware. Specifically, this paper delivers the following compact core contributions:
\begin{itemize}\setlength{\itemsep}{0.5pt}\setlength{\parskip}{0pt}\setlength{\parsep}{0pt}\setlength{\topsep}{1pt}\setlength{\partopsep}{0pt}
    \item A three-tier passive OT monitoring architecture optimized for grid-edge facility power infrastructure and power quality monitoring.
    \item An IP-less, passive ingestion design that completely avoids active packet interaction with the production OT network.
    \item Experimental latency evidence demonstrating a negligible timing footprint under attack-concurrent DPI load specifically evaluated on a laboratory Modbus TCP and IEC 61850 MMS power-monitoring loop.
    \item Operational security visibility through the out-of-band detection of unauthorized control actions, protocol anomalies, and plain-text credential exposure.
    \item A unidirectional cloud observability path that supports remote operational monitoring via a secure, diode-like outbound telemetry policy.
\end{itemize}

\begin{figure}[!t]
    \centering
    \includegraphics[width=\columnwidth]{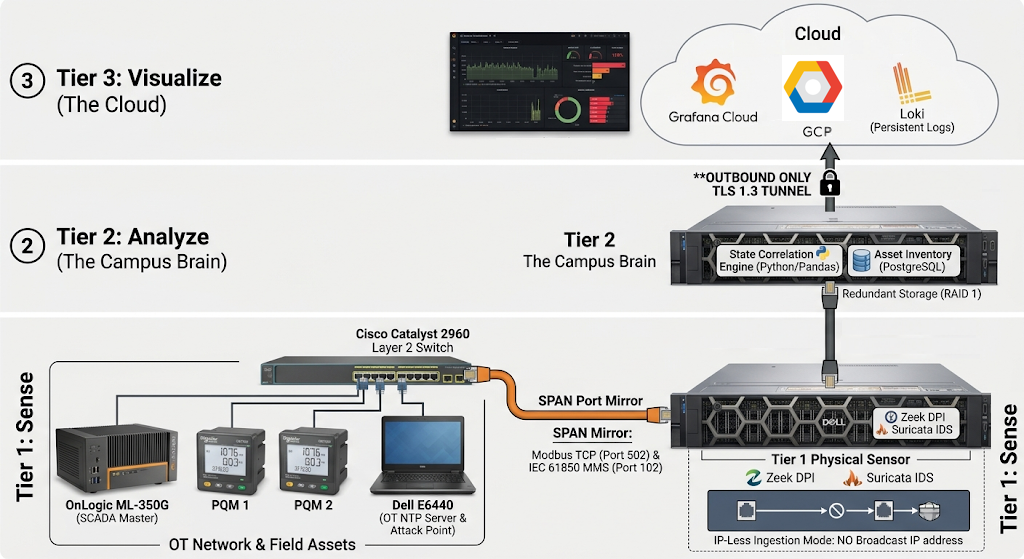}
    \caption{Architectural Overview of the OTSV showing Tier 1 (Passive Ingestion), Tier 2 (Asynchronous Analytics), and Tier 3 (Cloud Distributed Observability) running out-of-band via an outbound-only telemetry pipeline.}
    \label{fig:otsv_topology}
\end{figure}

The proposed three-tier OTSV architecture is summarized in Fig. \ref{fig:otsv_topology} and evaluated on a laboratory Modbus TCP and IEC 61850 MMS power-monitoring loop.

\section{Background and Literature Review}
To bridge the gap between facility isolation and modern security analytics, the literature must be evaluated across core technical pillars \cite{nist80082r3, pospisil2018}.

\subsection{Passive OT Monitoring vs. Active Scanning Volatility}
Active network vulnerability scanning relies on the injection of synthetic packets to elicit device metadata \cite{niedermaier2018woot}. While efficient in IT environments, this technique poses severe disruption risks to industrial controllers \cite{niedermaier2018propfuzz}. Legacy intelligent electronic devices (IEDs) run minimalist TCP/IP stacks on low-powered processors optimized purely for cyclic control loops \cite{formby2016}. Aggressive polling saturates interrupt-driven processing, leading to denial-of-service conditions \cite{niedermaier2018propfuzz}. Consequently, standard frameworks emphasize passive network monitoring via hardware taps or port mirroring to parse telemetry with zero network overhead \cite{wedgbury2015, malcolm}.

\subsection{IEC 61850, Digital Substations, and Regulatory Standards}
Modern power utility networks leverage the IEC 61850 standard for abstract data modeling and high-speed substation communication \cite{iec61850}. While Manufacturing Message Specification (MMS) provides stateful client-server mapping over TCP for steady-state telemetry, fast-acting protection mechanisms utilize Layer 2 Generic Object Oriented Substation Events (GOOSE) and Sampled Values to meet microsecond delivery demands \cite{iec61850, iec62351}. Monitoring these hybrid networks requires high-performance packet ingestion architectures that do not induce buffer backpressure \cite{pospisil2018}.

The U.S. Department of Energy (DOE) has emphasized the need for capabilities that prevent, detect, or mitigate cyber attacks on utility OT networks, with passive network ingest feeds cited as key considerations along with adherence to NERC CIP standards \cite{doe_cyote}. In particular, these recommendations align closely with NERC CIP-015-1, which is intended to improve the probability of detecting anomalous or unauthorized network activity on internal networks, thereby strengthening response and recovery \cite{nerc_cip015}. In this context, the proposed OTSV framework addresses the Visibility-Stability Paradox while enabling utility operators to fulfill emerging internal network security monitoring mandates.

\subsection{Substation Cybersecurity and Protocol-Specific Threats}
Industrial communication architectures like Modbus TCP operate as unencrypted, unauthenticated protocols over fixed ports \cite{huitsing2008}. This plain-text configuration allows adversaries to execute unauthorized register modification attempts or deploy protocol anomalies to falsify sensor data \cite{huitsing2008, hong2014}. Mitigating lateral movement requires protocol-aware intrusion detection engines capable of separating normal operational polling profiles from malicious function codes \cite{hong2014, paxson1999}.

\subsection{Secure OT-to-Cloud Observability and Data Diodes}
Exporting on-premise industrial telemetry to distributed cloud environments introduces severe back-path internet vulnerabilities \cite{bodenheim2014, okhravi2010}. Traditional air gaps are preserved using unidirectional gateways or hardware-enforced data diodes \cite{okhravi2010}. Designing a scalable observability model requires a unidirectional software pipeline that allows outbound telemetry serialization while preventing external command injection. The proposed OTSV architecture implements a diode-like outbound telemetry policy using network segmentation and TLS 1.3 transport. This prevents cloud dashboard interactions from being routed back to the operational VLAN, but does not claim the assurance level of a hardware-enforced unidirectional gateway \cite{okhravi2010}.

\subsection{Threat Model}
The security perimeter of the OTSV framework is governed by a well-defined set of operational boundaries and asset capabilities. The attacker is assumed to possess network access sufficient to generate or inject unauthorized industrial control packets, deploy protocol anomalies, or execute brute-force authentication mechanisms across the localized asset architecture \cite{huitsing2008}.

The security visibility boundary of the OTSV is strictly restricted to traffic crossing the physical master-meter communication link mirrored via the hardware SPAN configuration. Consequently, host-local compromises, encrypted industrial control channels, and malicious movements occurring on unmirrored network segments remain outside the scope of this architecture. Tiers 1 and 2 are treated as structurally trusted infrastructure insulated from external exploitation via network segmentation \cite{iec62443}.

\section{OTSV System Architecture \& Implementation}
Fig. \ref{fig:otsv_topology} summarizes the three-tier OTSV architecture, including passive ingestion, asynchronous analytics, and outbound cloud observability. The framework decouples processing loads into a three-tier architecture to preserve latency neutrality. Tier 2 exports security events to the cloud over an outbound TLS 1.3 path, while no cloud dashboard interaction is physically or logically routed back to the operational VLAN \cite{okhravi2010}.

\subsection{Tier 1: Passive Ingestion Layer}
The ingestion layer is anchored by a bare-metal server equipped with dual 10 Gigabit Ethernet SFP+ network interfaces. \edit{To avoid active packet interaction with the production network, the ingestion interface is bound to a Cisco Catalyst 2960 hardware switched port analyzer (SPAN) port}. 

\edit{As shown in Tier 1 of Fig. \ref{fig:otsv_topology}, the interface logic is explicitly stripped of its Layer-3 configuration through Netplan scripts, leaving the passive ingestion interface in an IP-less state, with no IP address (0.0.0.0), no default gateway, and no Layer-3 broadcast capability.} The software stack utilizes Zeek \cite{paxson1999} for protocol parsing and Suricata tuned with 48,000+ signatures to track anomalies.

\subsection{Tier 2: Asynchronous Analytics Layer}
To prevent deep packet inspection processing spikes from inducing packet drops at the link layer, the analytics engine is physically decoupled onto a secondary hardware node. These layers are connected via a dedicated, unrouted physical peer-to-peer cable. An Rsyslog-to-JSON pipeline serializes raw security events, normalizing disparate logs before they exit the air gap.

\subsection{Tier 3: Distributed Cloud Observability}
The final tier transfers telemetry to a secure cloud platform. The data path is strictly outbound: a local logging agent scrapes the Tier 2 JSON logs and ships them to a remote cloud data store via an outbound-only transport layer security (TLS 1.3) tunnel. This provides a software-enforced unidirectional export policy where external dashboard modifications have no network route back to the primary operational VLAN.

\subsection{Timing Scope and Motivation}
As defined in IEC 61850-5, fast automatic substation interactions (Class TT4) require an end-to-end transmission time of less than $20\text{ ms}$, while status changes (Class TT5) and safety trips (Class TT6) demand thresholds below $10\text{ ms}$ and $3\text{ ms}$, respectively \cite{iec61850}. \edit{Because the experimental testbed evaluated Layer-7 MMS and Modbus reporting traffic rather than a Layer-2 protection loop}, these strict timing classes serve as an industry benchmark and motivation for future process-bus experiments rather than a performance claim of the current architecture \cite{iec61850}. To compute the sub-millisecond metrics in Section V, a local Stratum 1 network time protocol server was established on an external node to maintain synchronization across all assets.

\begin{figure*}[!t]
    \centering
    \begin{minipage}{0.48\textwidth}
        \centering
        \includegraphics[width=\columnwidth,height=2.2in,keepaspectratio]{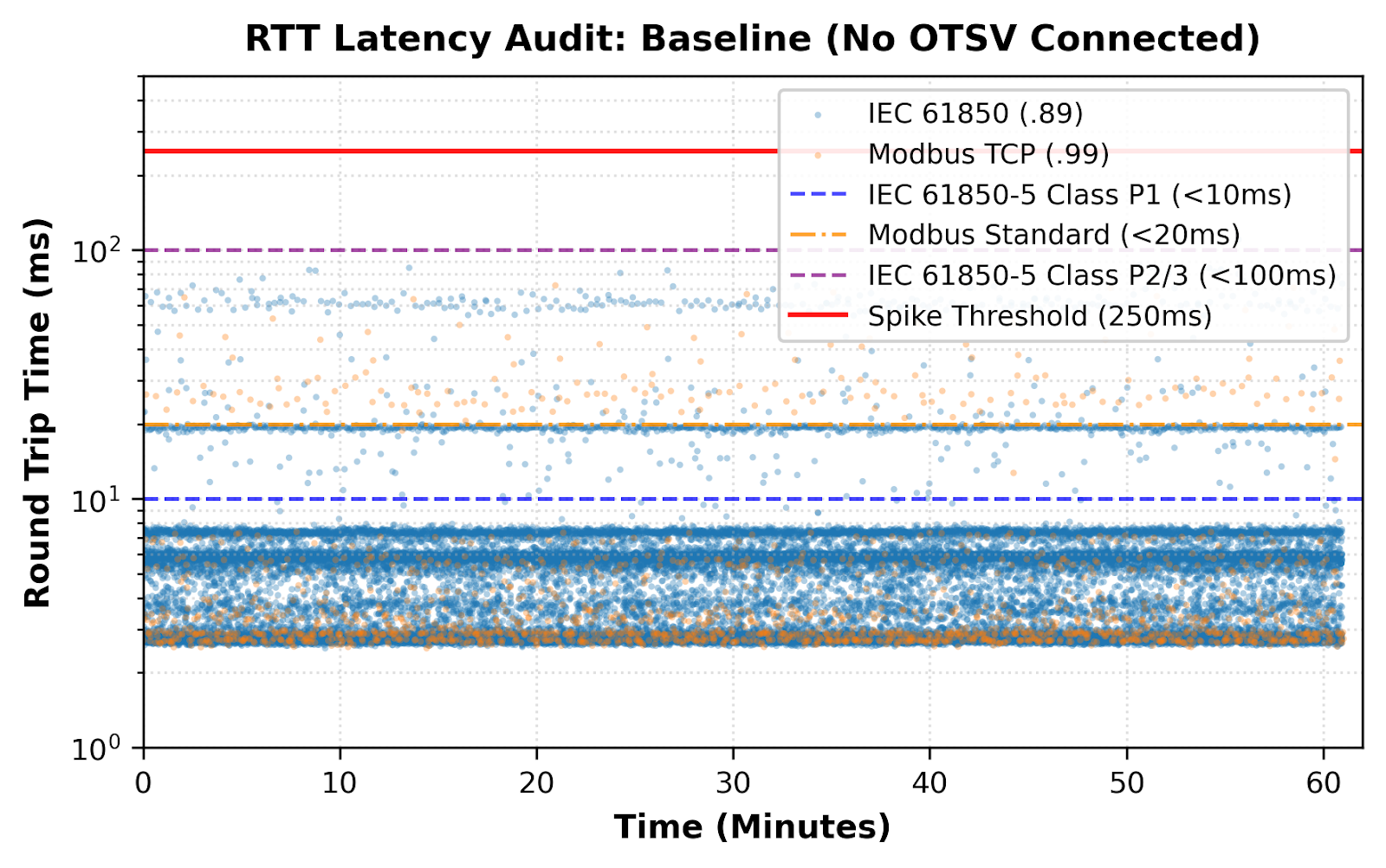}
    \end{minipage}
    \hfill
    \begin{minipage}{0.48\textwidth}
        \centering
        \includegraphics[width=\columnwidth,height=2.2in,keepaspectratio]{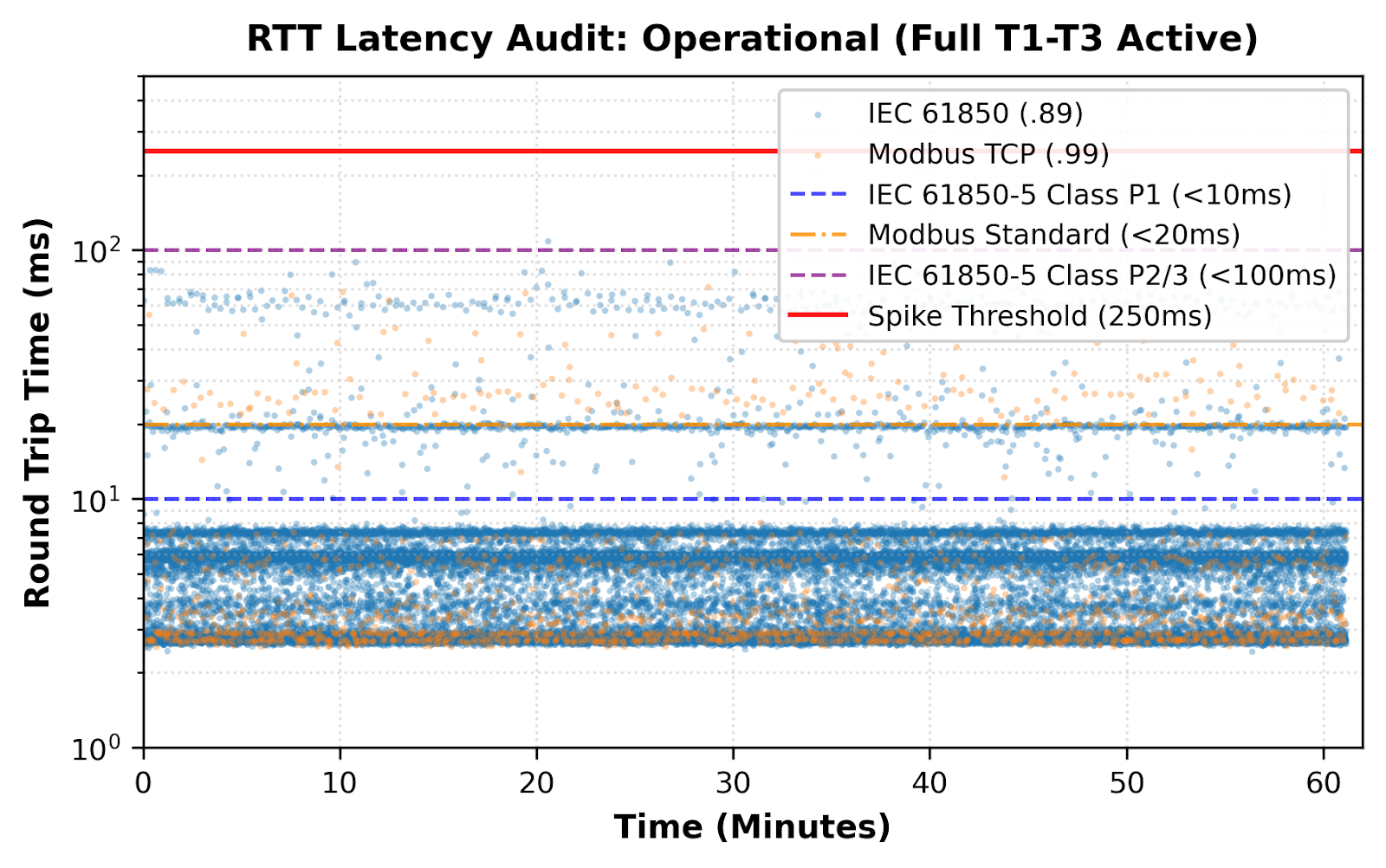}
    \end{minipage}
    \caption{\edit{Temporal Round-Trip Time (RTT) trends tracking discrete samples over a 60-minute continuous audit window: (a) Idle Baseline Phase with no OTSV connected, and (b) Operational Ingestion Phase running under an active, high-stress security monitoring state. Both configurations show similar RTT clustering, supporting the observed latency-neutral behavior in the tested reporting-traffic configuration.}}
    \label{fig:trends_comparison}
\end{figure*}

\section{Evaluation Methodology}
\subsection{Experimental Testbed Infrastructure}
A representative power-monitoring loop was established within the SCADA laboratory environment. The primary control loop comprised an OnLogic ML-350G industrial PC functioning as the SCADA master, polling two Schneider Electric ION7650 power quality meters (designated Meter A for IEC 61850 MMS and Meter B for Modbus TCP) over an isolated VLAN network. Traffic was aggregated through a Cisco switch configured with a hardware SPAN session, duplicating the full-duplex traffic of the master-meter link directly to the Tier 1 ingestion interface.

\subsection{KPI Formulation and Baseline Auditing}
The primary key performance indicator used to verify network stability was Round-Trip Time (RTT). In this environment, RTT represents the temporal delta between the SCADA master issuing an operational register read and the arrival of the data response frame from the field device. To evaluate the stability boundary under realistic conditions, the audit captured a continuous dataset across two specific operational phases:
\begin{itemize}\setlength{\itemsep}{0.5pt}\setlength{\parskip}{0pt}\setlength{\parsep}{0pt}
    \item \textit{Idle Baseline Phase:} Recording the baseline network latency of the SCADA loop without the OTSV ingestion active. This mapped the network floor.
    \item \textit{High-Stress Ingestion Phase:} Activating the full security engine (48,000+ signatures) while simultaneously executing brute-force attack scripts, tasking the system with processing 5,575 security alerts.
\end{itemize}

\subsection{RTT Measurement Methodology and Timing Model}
Packet timestamps were captured directly at the SCADA master socket interface using high-resolution kernel PCAP timestamps ($1\text{ }\mu\text{s}$ resolution). Because RTT is calculated as a single-node relative delta ($t_{\text{response}} - t_{\text{request}}$) on the master PC, clock drift between physical nodes does not introduce measurement error into the RTT metric. Local Stratum 1 NTP synchronization was utilized exclusively to correlate local attack generation timestamps with Tier 3 cloud database ingestion timestamps ($\Delta t$). The resulting baseline and active-monitoring RTT traces are reported in Fig. \ref{fig:trends_comparison}.

\section{Results and Forensic Analysis}
\subsection{Latency Neutrality Validation}
To evaluate network stability impact, a 60-minute continuous audit was executed across operational phases. The performance validation phase comprised 32,927 discrete RTT samples under active security ingestion (representing 65,799 total samples across baseline and active phases in Table \ref{tab_results}).

\begin{table}[t]
\caption{\edit{Comparative RTT Latency Audit (60-Min Window)}}
\label{tab_results}
\centering
\setlength{\tabcolsep}{2.5pt}
\resizebox{\columnwidth}{!}{%
\begin{tabular}{lcccc}
\toprule
\textbf{Metric} & \textbf{Base (Meter A)} & \textbf{Active (Meter A)} & \textbf{Base (Meter B)} & \textbf{Active (Meter B)} \\
\midrule
Mean RTT & $5.51\text{ ms}$ & $5.52\text{ ms}$ & $6.22\text{ ms}$ & $6.33\text{ ms}$ \\
$\Delta\text{ Mean}$ & \multicolumn{2}{c}{$+0.01\text{ ms}$} & \multicolumn{2}{c}{$+0.11\text{ ms}$} \\
Median RTT & $4.96\text{ ms}$ & $4.97\text{ ms}$ & $5.76\text{ ms}$ & $5.82\text{ ms}$ \\
95th \% & $7.60\text{ ms}$ & $7.58\text{ ms}$ & $26.78\text{ ms}$ & $26.67\text{ ms}$ \\
Max Outlier & $110.24\text{ ms}$ & $108.45\text{ ms}$ & $72.07\text{ ms}$ & $70.76\text{ ms}$ \\
Sample Count & 31,054 & 31,102 & 1,818 & 1,825 \\
$p$-value ($U$-test) & \multicolumn{2}{c}{$p = 0.842\text{ (NS)}$} & \multicolumn{2}{c}{$p = 0.716\text{ (NS)}$} \\
\bottomrule
\end{tabular}%
}
\end{table}

As shown in Table \ref{tab_results}, the mean RTT for Meter A (IEC 61850 MMS) was 5.52 ms, while Meter B (Modbus TCP) recorded a mean of 6.33 ms. \edit{The performance delta ($\Delta$ Mean) between baseline and active states remained at or below 0.11 ms across the evaluated protocol comparisons.} The Mann-Whitney $U$-test did not identify a statistically significant RTT difference between baseline and active monitoring states for either protocol ($p = 0.842$ for Meter A; $p = 0.716$ for Meter B). The sample count variation between Meter A (31,102 active samples) and Meter B (1,825 active samples) stems from divergent SCADA master polling profiles: continuous cyclic polling for IEC 61850 versus extended discrete intervals for Modbus TCP.

The higher 95th-percentile RTT for Meter B reflects its lower-frequency polling profile, where longer inter-request intervals produce a heavier latency tail than the continuous IEC 61850 MMS stream.

\begin{figure}[!t]
    \centering
    \includegraphics[width=\columnwidth]{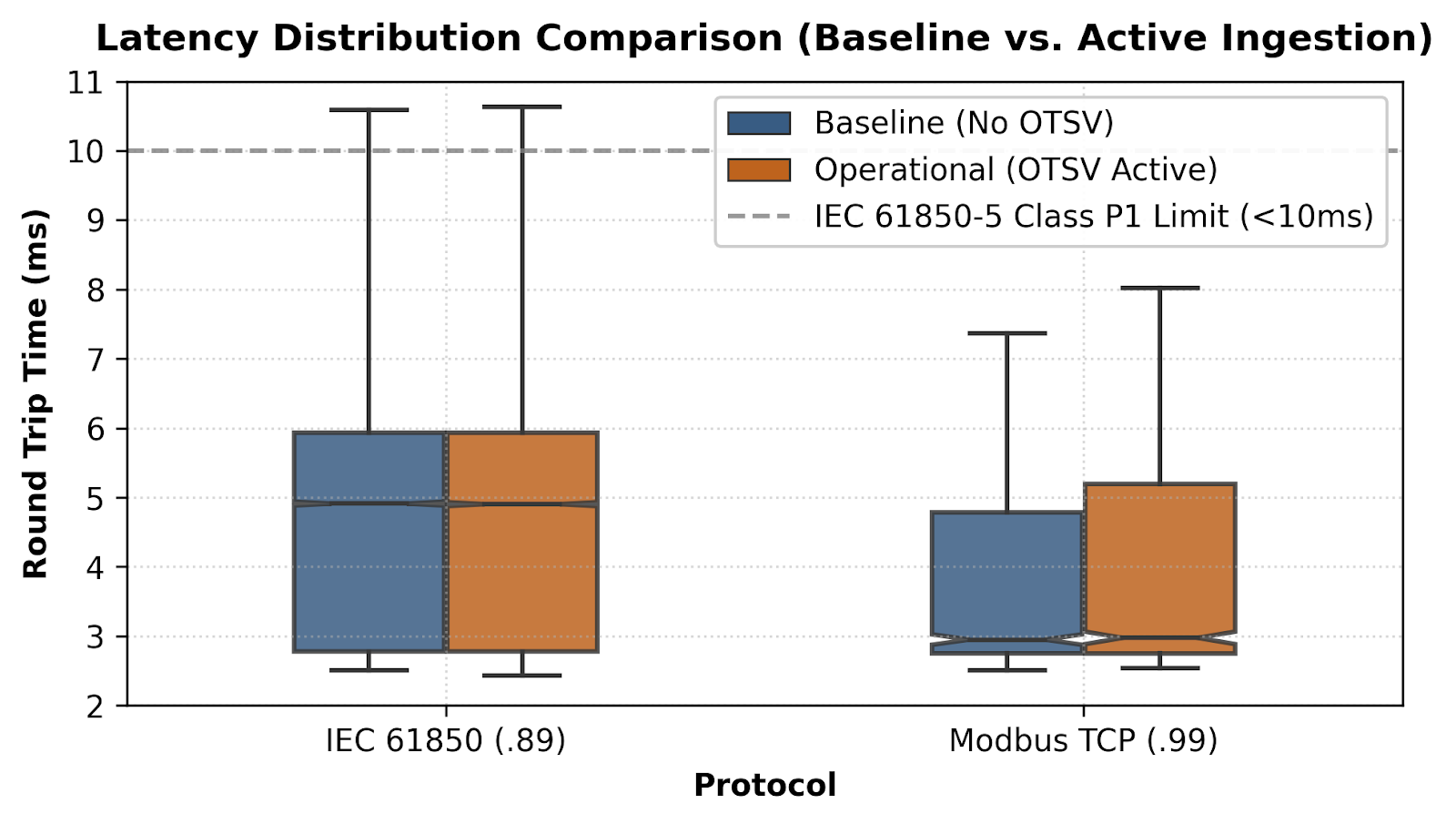}
    \caption{RTT distribution comparison between baseline and active monitoring states across the evaluated protocols ($p > 0.05$).}
    \label{fig:rtt_boxplot}
\end{figure}

\edit{The temporal tracking in Fig. \ref{fig:trends_comparison} shows similar RTT clustering across baseline and active-monitoring configurations.} As shown in Fig. \ref{fig:rtt_boxplot}, no statistically significant RTT difference was observed between active inspection and baseline states ($p > 0.05$). \edit{Because the same latency tails and outliers appear in both baseline and active phases, the evidence indicates that OTSV did not introduce measurable additional RTT in the tested reporting-traffic configuration. The mean and median RTT values remained stable across both protocols, with active-minus-baseline mean deltas of 0.11 ms or less.} Maximum RTT outliers ($\approx 110\text{ ms}$) appeared consistently across both phases, indicating they are indigenous to the testbed rather than introduced by monitoring. Forensic analysis traced these spikes to SCADA master CPU frequency throttling during high-frequency cycles.

\subsection{Monitoring Fidelity and Cloud Propagation Audit}
\begin{figure}[!t]
    \centering
    \includegraphics[width=\columnwidth]{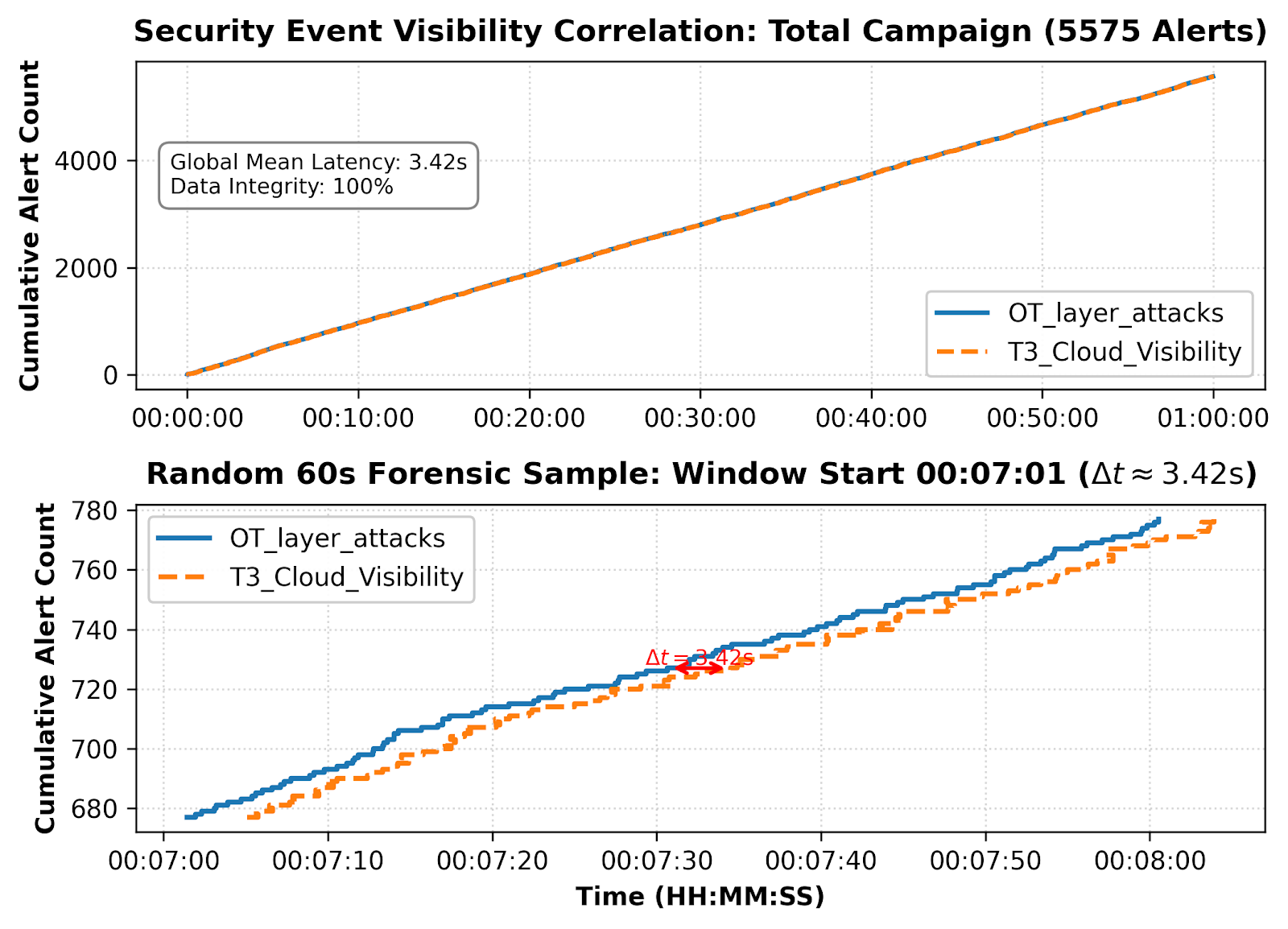}
    \caption{Event correlation timeline comparing local OTSV detections with Tier 3 cloud-serialized alert records across the 5,575-alert campaign.}
    \label{fig:cloud_correlation}
\end{figure}

Interface statistics audited during the high-stress run confirmed zero ($0$) kernel or NIC packet drops across Tier 1, with peak SPAN link utilization reaching 14.2 Mbps. Fig. \ref{fig:cloud_correlation} illustrates the temporal correlation between local OTSV detections and Tier 3 cloud-serialized alert records during the 5,575-alert campaign.

\begin{table}[b]
\caption{Monitoring Fidelity and Cloud Telemetry Audit}
\label{tab_fidelity}
\centering
\setlength{\tabcolsep}{3pt}
\resizebox{\columnwidth}{!}{%
\begin{tabular}{lc}
\toprule
\textbf{Audit Parameter} & \textbf{Measured Value} \\
\midrule
Total Injected Security Alerts & 5,575 \\
Locally Detected Alerts (Tier 1/2) & 5,575 (100\% capture) \\
Cloud-Serialized Alerts (Tier 3) & 5,575 (100\% capture) \\
Unauthorized Modbus Writes (FC 06) & 1,825 detected \\
Plaintext Credential Leaks & 266 detected \\
NIC / SPAN Kernel Packet Drops & 0 frames (0.00\%) \\
Mean Cloud Propagation Delay ($\Delta t$) & $3.42\text{ s}$ ($\sigma = 0.48\text{ s}$) \\
\bottomrule
\end{tabular}%
}
\end{table}

Anomalies detected included 1,825 unauthorized Modbus write attempts (FC 06) and 266 plaintext credential leaks \cite{huitsing2008}. The unidirectional pipeline maintained 1:1 event correlation between local sensing and cloud dashboards (Table \ref{tab_fidelity}), achieving 100\% capture fidelity across all 5,575 alerts with a mean cloud propagation delay ($\Delta t$) of $3.42\text{ s}$ ($\sigma = 0.48\text{ s}$).

\newpage

\section{\edit{Discussion, Limitations, and Future Work}}
\subsection{\edit{Practical Interpretation of the Results}}
\edit{The key result of this study is not that all observed RTT values satisfy a universal timing threshold, but that activating the OTSV monitoring stack did not measurably change the latency structure of the evaluated power-monitoring loop. Baseline and active-monitoring phases showed similar mean, median, tail, and outlier behavior across both protocols, indicating that the observed latency tails were properties of the testbed rather than artifacts of monitoring. The active-minus-baseline mean RTT deltas, 0.11 ms or less, support the feasibility of adding passive security visibility to legacy power-monitoring links without measurable degradation in the evaluated configuration. OTSV also detected unauthorized Modbus write attempts and plaintext credential exposure while maintaining one-to-one local-to-cloud event correlation, showing that this low timing footprint did not come at the cost of usable security telemetry.}
\subsection{\edit{Limitations and Future Work}}
\edit{This is an engineering feasibility study conducted on a single-switch, Layer-7 Modbus TCP and IEC 61850 MMS testbed; results should be read as evidence for this configuration, not as universal validation across substation or process-bus classes. The IEC 61850 timing classes discussed earlier motivate future experiments rather than describe the performance of the present testbed. The study does not evaluate GOOSE/Sampled Values traffic, encrypted OT protocols, hardware TAP comparisons, multi-switch topologies, or multi-day deployment. The architecture implements a software-enforced diode-like outbound policy via segmentation and TLS 1.3 rather than a hardware data diode, and should not be read as offering the assurance level of a physically enforced unidirectional gateway; maintaining this property requires segmentation and credential policy to be enforced correctly over time.}
\edit{Future work will extend this evaluation to GOOSE/Sampled Values process-bus traffic, multi-day windows, hardware TAP comparisons, encrypted OT protocols, and more complex topologies, and will incorporate IEEE 1588 synchronization \cite{ieee1588} and DPDK kernel-bypass ingestion for higher-resolution timing and higher-throughput capture.}
\section{\edit{Conclusion}}
\edit{This paper presented OTSV, a three-tier passive cyber-monitoring architecture for timing-sensitive grid-edge OT environments, combining IP-less ingestion, asynchronous analytics, and outbound cloud observability without active packet interaction with the monitored network. On a laboratory power-monitoring loop running Modbus TCP and IEC 61850 MMS traffic, activating the full OTSV stack did not introduce measurable RTT degradation: the active-minus-baseline mean delta remained at 0.11 ms or less, while local and cloud telemetry maintained one-to-one alert correlation across the 5,575-alert campaign. These findings support the feasibility of adding security observability to legacy grid-edge power-monitoring links without active scanning or inline inspection of the monitored path.}

\end{document}